\documentclass[sigconf]{acmart}

\AtBeginDocument{%
  \providecommand\BibTeX{{%
    \normalfont B\kern-0.5em{\scshape i\kern-0.25em b}\kern-0.8em\TeX}}}

\usepackage{booktabs} 
\usepackage{multirow}
\usepackage{xcolor}
\usepackage{algorithmic}
\usepackage{mmstyle}
\usepackage[ruled]{algorithm2e} 

\newif\ifshownotes

\ifdefined\MakeWithNotes
\shownotestrue
\fi
\ifdefined\MakeWithoutNotes
\shownotesfalse
\fi

\ifshownotes

\newcommand{\colornote}[3]{{\color{#1}\bf{#2: #3}\normalfont}}
\newcommand{\colornoteTwo}[3]{{\color{#1}\bf{#3}\normalfont}}
\newcommand{\colornoteThree}[2]{{\color{#1}\bf{#2}\normalfont}}      
\else
\newcommand{\colornote}[3]{}
\newcommand{\colornoteTwo}[3]{}
\newcommand{\colornoteThree}[2]{}      
\fi

\newcommand{\model}{ScriptViz\xspace}

\SetAlFnt{\small}
\SetAlCapFnt{\small}
\SetAlCapNameFnt{\small}
\SetAlCapHSkip{0pt}

\copyrightyear{2024}
\acmYear{2024}
\acmDOI{XXXXXXX.XXXXXXX}
\acmConference[UIST '24]{The ACM Symposium on User Interface Software and
Technology}{Oct. 13--16,
  2024}{Pittsburgh, PA}

\begin{document}
\title{ScriptViz: A Visualization Tool to Aid Scriptwriting based on a Large Movie Database}


\author{Anyi Rao}
\affiliation{%
  \institution{Stanford University}
   \country{}
}
\author{Jean-Peïc Chou}
\affiliation{%
  \institution{Stanford University}
     \country{}
}
\author{Maneesh Agrawala}
\affiliation{%
  \institution{Stanford University}
     \country{}
}



\begin{abstract}
Scriptwriters usually rely on their mental visualization to create a vivid story
by using their imagination to see, feel, and experience the scenes they are writing.
Besides mental visualization, they often refer to existing images or scenes in movies and analyze the visual elements to create a certain mood or atmosphere.  
In this paper, we develop \model to provide external visualization based on a large movie database for the screenwriting process. 
It retrieves reference visuals on the fly based on scripts’ text and dialogue from a large movie database.
The tool provides two types of control on visual elements that enable writers to  1) see exactly what they want with fixed visual elements and 2) see variances in uncertain elements.
User evaluation among 15 scriptwriters shows that \model is able to present scriptwriters with consistent yet diverse visual possibilities, aligning closely with their scripts and helping their creation.
\end{abstract}

%
%


%
%


\begin{teaserfigure}
	\includegraphics[width=0.99\linewidth]{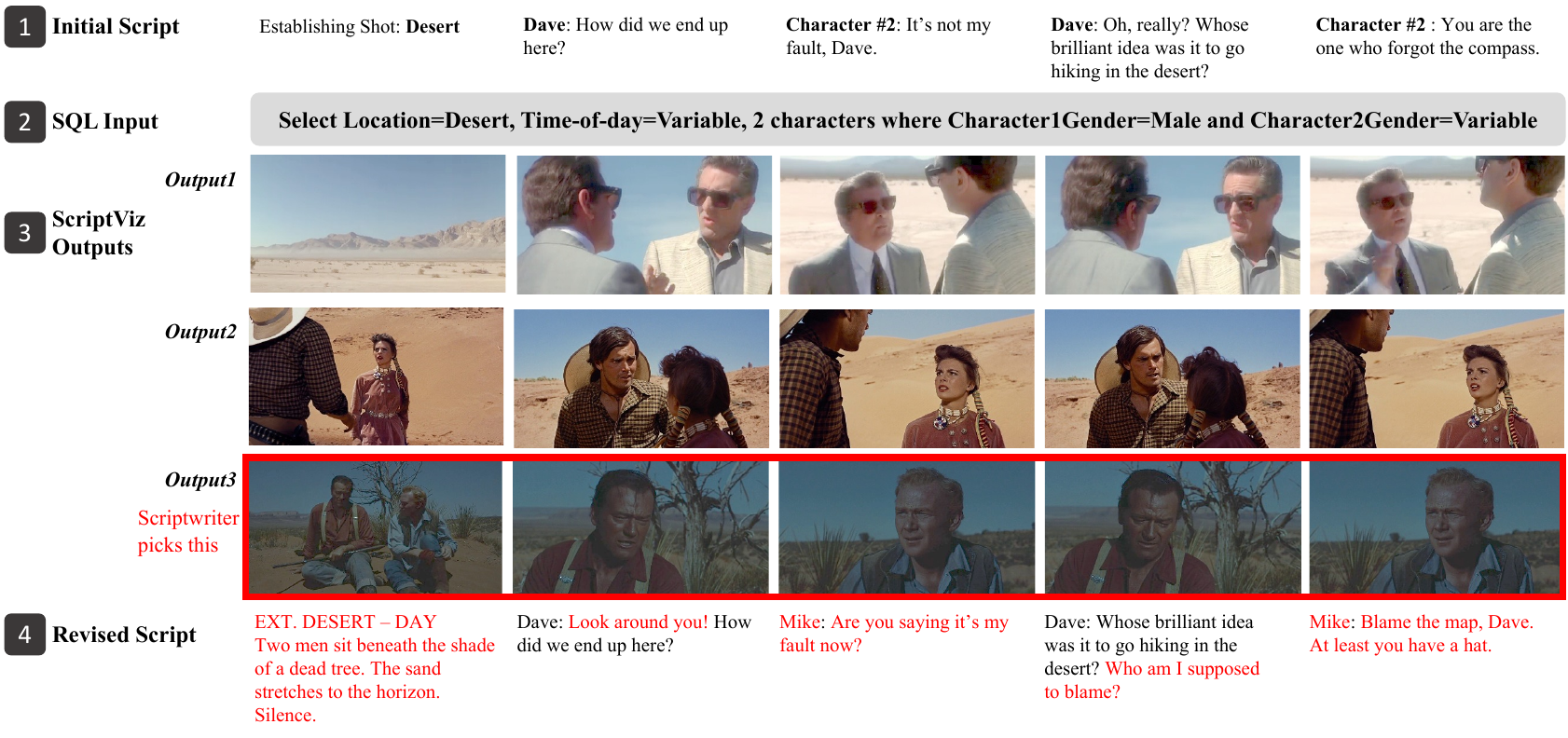}
	\vspace{-15pt}
	\caption{A scriptwriter writes a main character ``Dave", who is irritable and impulsive, trapped in a desert. 
    She has a few sentences in her mind as shown in the top. 
    She wants to choose an appropriate setting on time-of-day and the other character's gender at the beginning of her creation.
    By setting these as variant attributes,
    she sees different visualizations in three rows and prefers the 3rd one.
    The night view she sees better reflects characters' bad situations as they are tired for a whole day. 
    The dead tree in the background strengthens this tension and also changes her script by adding details into it (bolded red text) at the bottom.
    These additions create a better main character by enhancing Dave's personality. After seeing the visualization, it helps to add details that better reflect characters' desperate situation. 
    The interface is shown in the supplementary video.}
	\label{fig:teaser}
	\vspace{-0pt}
\end{teaserfigure}

\maketitle

\section{Introduction}


Scriptwriting is a creative process that goes beyond writing text; it is the process of developing a blueprint that will be turned into a visual story.
While scripts inherently lack visual elements, they must effectively convey visual ideas, which are often as integral as the narrative itself in the story's final cinematic form\,\cite{field2005screenplay, mckee1997story}.
Scriptwriters often mentally visualize scenes as they write them
to envision how visual attributes of its setting 
and characters 
might affect its tone, atmosphere, and character interactions.  
They use mental visualizations to adjust the script to produce a more visual blueprint for the film.
For example, imagining a suspenseful scene taking place at night in a rain-soaked alley between two middle-aged women could help the writer 
add elements to the scene descriptions and dialogue that evoke film noir.

One common method amongst scriptwriters is to ground their scripts using visual references from existing movies.
For instance, the scriptwriters for
\emph{The Silence of the Lambs (1991)} used the movie \emph{The Manhunter (1986)} as a visual reference when writing the conversion between Hannibal Lecter and Clarice Starling~\cite{silence}.
Similarly, the scriptwriters for 
\emph{The Dark Knight (2008)}, used the movie \emph{Heat (1995)} as a visual reference when writing the dialogue scene between Bruce Wayne and Alfred Pennyworth\,\cite{dark}.

Although storyborads and moodboards may look similar to our tool, they serve totally different goals.
Storyboards typically visualize camera angles and character staging for each shot. 
They are usually used to guide the filming process, but not so much during scriptwriting. 
Moodboards are often constructed at the same time as writing the script, but they do not visualize the scenes playing out. 
Our script visualizations fit between these two as they are designed to aid scriptwriting but at the level of visual attributes of the scene rather than camera angles and staging.

In this work, we introduce ScriptViz, a tool for scriptwriters to visualize a scene as they are writing it using reference imagery from existing feature films.
It retrieves sequences from a pre-processed database of existing movies that match the writer's script and inputs. 
\model uses dialogues-aligned visualization and selects frames for each sequence to build a new sequence of illustrations corresponding to the writer's evolving script. 
Our database preprocessing method can successfully retrieve clear frames of establishing shots and characters, offering a tangible cinematic representation of their script to rely on for mental play-through or visual ideation.

We pinpoint the visual elements of settings and characters as being instrumental for scriptwriters and support these elements being set as fixed and variable: 
1) fixed attributes that scriptwriters have definitive ideas about. For instance, they might envision a specific setting, such as a bedroom, that is crucial to the scene's atmosphere and narrative; 
2) variable attributes where scriptwriters seek flexibility and variety, allowing them to explore different possibilities. An example is a character's age, like imagining the character Sarah at various ages to see how it changes the scene's dynamics.

More specifically, given a partial script and 
a specification of fixed and variable visual attributes, 
ScriptViz identifies scenes from existing films that match the specifications. 
For each such scene, it retrieves a set of keyframes from the original scene and identifies a keyframe that best presents an establishing shot of the setting. 
It also assigns characters from the input script to recurring faces in the retrieved keyframes and then chooses a keyframe for each line of dialogue in the script while ensuring that the speaking character is visible in the frame.

For instance, 
in a real scriptwriter usage case, as shown in Figure~\ref{fig:teaser},
the visualization helps to add details that better reflect the character's desperate situation. 
We further demonstrate the effectiveness of our system by showing the results of 3 practical scriptwriting cases in Sec.~\ref{sec:results}.
The user evaluation and comparison with three baselines among 15 scriptwriters in Sec.~\ref{sec:experiment}
further shows that the tool is able to aid the scriptwriting process.

Our work makes two main contributions. 
(1) We provide a tool that lets scriptwriters visualize their scripts as they are writing them to support brainstorming and planning via retrieving existing movie frames and changing semantic attributes to ensure strong consistency and rich variance. 
(2) We identify visual attributes of settings and characters that scriptwriters find useful for imagining their scripts. 
And our user study shows that letting scriptwriters place these attributes into two categories, fixed and variable, enables better brainstorming and planning.
\section{Related Work}
\label{sec:related}

\vspace{0.5em}
\noindent\textbf{Storyboards and moodboards.}
Storyboards and moodboards are traditional tools integral to the filmmaking process, used to pre-visualize story ideas and scripts~\cite{chen2024cinepregen}. Storyboards, visual sequences of illustrations, outline each shot to guide the filmmaking process. They detail camera angles and character positions, assisting filmmakers in planning and visualizing shot composition post-screenwriting. Conversely, moodboards, which are collections of visuals and text, are employed to convey the overall tone and mood before the scriptwriting process begins. Our script visualization tool is designed to aid scriptwriters in crafting better stories by providing visualization support during the writing process. 

Researchers have explored various methods to assist storyboard and moodboard visualization.
\citet{goldman2006schematic} visualized short videos in a single static image.
\citet{rao2023dynamic} visualized a storyboard with existing 3D character and motion assets in a 3D virtual environment.
Moodboard tools such as \cite{milanote} allow creators to compile and organize visual references and color schemes,
thereby establishing a cohesive aesthetic vision for their projects~\cite{koch2020semanticcollage}. 
\citet{chou2023talestream} used tropes as an intermediate representation of stories to approach story ideation.
While storyboards provide visuals for movement guidance for crew members, and moodboards convey the overall style and feel of the project,
\model focuses on the visualization of a scene in scripts to help scriptwriting.

\vspace{0.5em}
\noindent\textbf{Visualizing stories.}
Researchers have used image generative models~\cite{creswell2018generative} to produce images for story visualization, which is aimed at helping viewers see visual imagery related to a story, such as StoryGAN~\cite{li2019storygan} and DUCOStoryGAN~\cite{maharana2021integrating}.
StoryDALL-E~\cite{maharana2022storydall} further improves its performance with a pre-trained transformer by leveraging latent knowledge from large-scale datasets.
\citet{gong2023interactive} 
added character images and text as conditions to text-to-image models to increase identity consistency~\cite{ye2023ip-adapter} and text-visual alignment in story visualization. 
\citet{rahman2023make} proposed a novel autoregressive
diffusion-based framework with a visual memory module
that implicitly captures the actor and background context across the generated frames.
\citet{pan2024synthesizing} proposed a latent diffusion model
auto-regressively conditioned on history captions and generated images.
The goal of story visualization is different from our script visualization which helps scriptwriters. 
A difficulty of using all of these generation-based methods for script visualization is that even when they are designed to produce consistent character identities or styles across multiple images they cannot guarantee such consistency. In contrast, our approach guarantees such consistency by using reference imagery from existing films while we also provide some consistent post edits in the form of setting changes (\eg weather), and character changes (\eg facial expression change) in the supplementary.

Research has also been dedicated to creating visual summaries from the films themselves.
\citet{xiong2019graph} proposed a graph-based method to match movie segments and synopsis paragraphs on MovieNet~\cite{huang2020movienet}.  
\citet{gu2023tevis} presented a VQ-Trans model to compute a joint embedding space for images from existing movies and the query synopsis.
Our method focuses on repurposing existing movie frames for the writing of a new script. 
Compared to existing movie frame retrieval methods that compute the similarity of text sentences and visual frames, 
our approach focuses on the accuracy of specific visual attributes, such as location and character identity, and offers the scriptwriters control over these attributes to help them imagine new scripts.

\vspace{0.5em}
\noindent\textbf{Story writing support systems.}
The domain of story writing support has received considerable attention, evolving significantly to encompass a variety of specialized systems designed to assist writers. Traditionally, commercial software like FinalDraft \cite{finaldraft}, Dramatica \cite{dramatica}, and Plottr \cite{plottr} has provided structured frameworks to help writers organize their writing, plots, and narratives. The analysis of the way stories are crafted and function has enabled the development of autonomous story generation, whether through computational planning \cite{lebowitz1984creating, meehan1977talespin, riedl2010narrative}, character-based simulations \cite{cavazza2001characters}, or case-based reasoning \cite{perez2001mexica, gervas2005story, riedl2009vignette, riedl2010narrative}. The recent advancement in language models has ushered in a new era of intelligent writing assistants \cite{lee2024design}. These tools, capable of generating textual propositions from keywords \cite{fan2018hierarchical, ippolito2019unsupervised, sun2021iga, xu2020megatroncntrl}, engaging users in conversation \cite{duval2021breaking, lee2022coauthor}, or interpreting multi-modal inputs \cite{chung2022talebrush}, have garnered considerable attention. Their applications and effectiveness in supporting story writing have been extensively explored \cite{clark2018creative, calderwood2020how, yuan2022wordcraft, biermann2022from, mirowski2023cowriting}, with a myriad of intelligent writing tools now readily available \cite{sudowrite, aidungeon}. The capacity of large language models has also been used to generate feedback and annotations, enriching the writing process with additional insights. Beyond textual assistance, support systems have also employed non-textual elements, including the use of tropes \cite{chou2023talestream} and game cards \cite{akoury2020storium} to stimulate creativity and narrative development. The integration of visuals into the creative process has been shown to enhance storytelling tasks \cite{ali2021telling}. Particularly relevant to our research is the VWP dataset that features sequences of movie shots aligned with stories collected via crowdsourcing. This dataset powers a character-based story generation model focused on visual storytelling tasks \cite{hong2023visual}.

In contrast to these approaches, our work with ScriptViz introduces a novel dimension by leveraging cinematic visualizations to support scriptwriters specifically. Unlike systems that primarily focus on narrative structure, generation, or inspiration, ScriptViz is designed to enhance mental visualization of scenes, offering writers tangible cinematic representations that facilitate the development of the visual potential of their scripts.
\section{System Design}
\label{sec:system}

At the heart of every script lies the potential to transform words into vivid, cinematic experiences. With \model, our primary goal is to encourage scriptwriters to engage with the visual elements of their storytelling. The design of our tool is specifically tailored to support scriptwriters in both the exploration and refinement of their work's visual dimensions and to understand their implications on the overarching story. The design of our system is motivated by common practices as well as film theory.

\begin{figure*}[!t]    
    \includegraphics[width=0.99\linewidth]{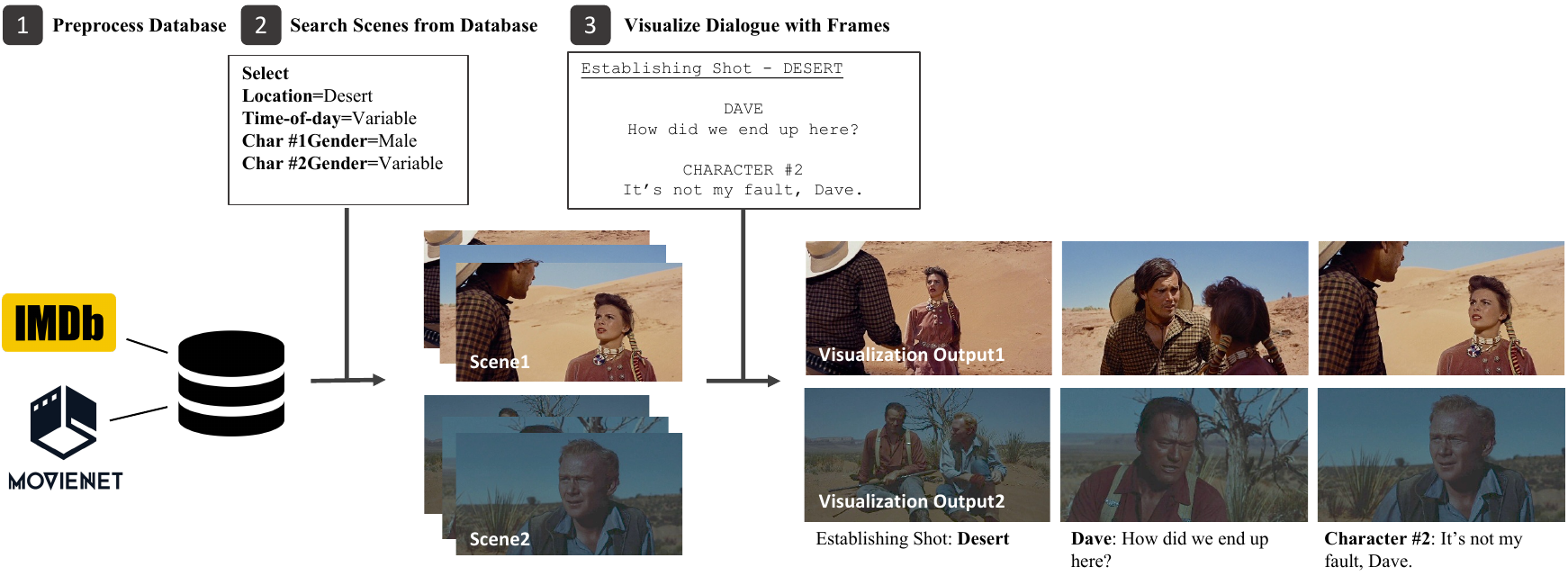}
    \vspace{-15pt}
    \caption{The pipeline of \model. It first preprocesses the database and retrieves scenes according to user's input.
    It is based on two types of control on visual attributes that enable writers to  1) see exactly what they want with fixed visual elements and 2) see variances in uncertain elements.
    It then retrieves images for each script’s dialogue sequences.
    }
    \label{fig:pipeline}
    \vspace{-5pt}
\end{figure*}

\vspace{2pt}
\noindent\textbf{Leveraging existing movies.}
Building upon traditional visualization practices like moodboards and storyboards, \model leverages existing movie footage. On the one hand, it selects sequences that capture the desired attributes as inspiration, akin to how scriptwriters curate visual references on moodboards. On the other hand, drawing upon storyboards, \model suggests possible outline visualizations of the narrative flow with sequences of shots. It achieves this by selecting and rearranging frames within each retrieved sequence to generate retargeted sequences of shots that align with the script.

This approach recontextualizes scenes from existing movies and is informed by montage theory, which posits that the arrangement of shots within a sequence shapes narrative meaning and can thus support different narrative purposes \cite{eisenstein1949film}. Besides, the use of a single sequence for each visualization ensures visual and narrative consistency, capitalizing on the continuity rules followed in most movie sequences \cite{bordwell2008film}. Drawing from a vast reservoir of well-known films guarantees rich, organic, and diverse illustrations characterized by well-composed frames and compelling character performances.

\vspace{2pt}
\noindent\textbf{Dialogue-aligned visualization.}
To engage scriptwriters with the visual aspect of their story and help them play out the scene more naturally, it's crucial that the retargeted sequences visually align with the input script. This process requires careful consideration of what to depict and when to depict it along the script, maintaining coherence and relevance throughout the sequence.

Given that conversations most often drive the progression of film sequences \cite{field2005screenplay} and correspond to over half the shots in most movies
\cite{cutting2015shot}, \model associates each line of dialogue with a corresponding frame featuring the relevant character. This approach mirrors classic cinematographic techniques, particularly the use of shots and reverse-shots alternately showing the characters conversing on screen to simply convey dialogue-driven scenes.
This approach is a clear and immediate portrayal of how a conversation unfolds visually within a scene, providing scriptwriters with a coherent visual reference alongside the dialogue.


\vspace{2pt}
\noindent\textbf{Attributes control.}
In \model, we offer controls over \emph{fixed} and \emph{variable} attributes to strike a balance between the scriptwriter’s vision and the potential for creative exploration. \emph{Fixed attributes} ensure that the visualizations remain true to the writer's vision and the story's established elements. Illustrations aligned with the writer's ideas can help play out a scene. Conversely, \emph{variable attributes} open the door to different possibilities, encouraging writers to explore various scenarios and contexts. 
An attribute that is specified as \emph{variable} or not specified is considered as \emph{variable}. 
For such attributes, \model retrieves sequences with as much diversity as possible to generate new perspectives and ideas that can be picked up to develop and refine the script. This distinction aims to invite the user to specify essential visual attributes for their scene.

We streamline the visualization controls by focusing on two primary visual storytelling attributes~\cite{eisenstein1949film,field2005screenplay} and additionally let the scriptwriters use their own descriptors. 

\vspace{0.5em}
\noindent\textit{Settings.}
The settings, which correspond to the location and the time of day, are not only an essential visual backdrop but also an active participant in the narrative, influencing mood and thematic attributes~\cite{mckee1997story}. Each setting brings its unique narrative connotations, visual language, and energy. \model allows scriptwriters to select settings that resonate with their story's atmosphere and themes to visually cohere with their writing and invite scriptwriters to specify this key visual attribute for each of their scenes.

\vspace{0.5em}
\noindent\textit{Character attributes.}
Visualizing characters can aid in developing their personalities and interactions with their environment and other characters. Character attributes such as gender and age contribute significantly to the definition of characters. By offering control over character attributes, ScriptViz lets writers develop and describe their characters visually, aligning with their envisioned narrative roles and interactions.

\section{Method}
\label{sec:methodology}

We develop \model according to the above design ideas, as shown in Figure~\ref{fig:pipeline}.
It contains three parts: 
(1) preprocess on the database with scene, shot, and frame-level attributes;
(2) find the corresponding scenes in the database given a request for fixed and variable attributes;
(3) choose an establishing frame and frames for
each line of dialogue.

\begin{figure*}[!t]    
    \includegraphics[width=0.99\linewidth]{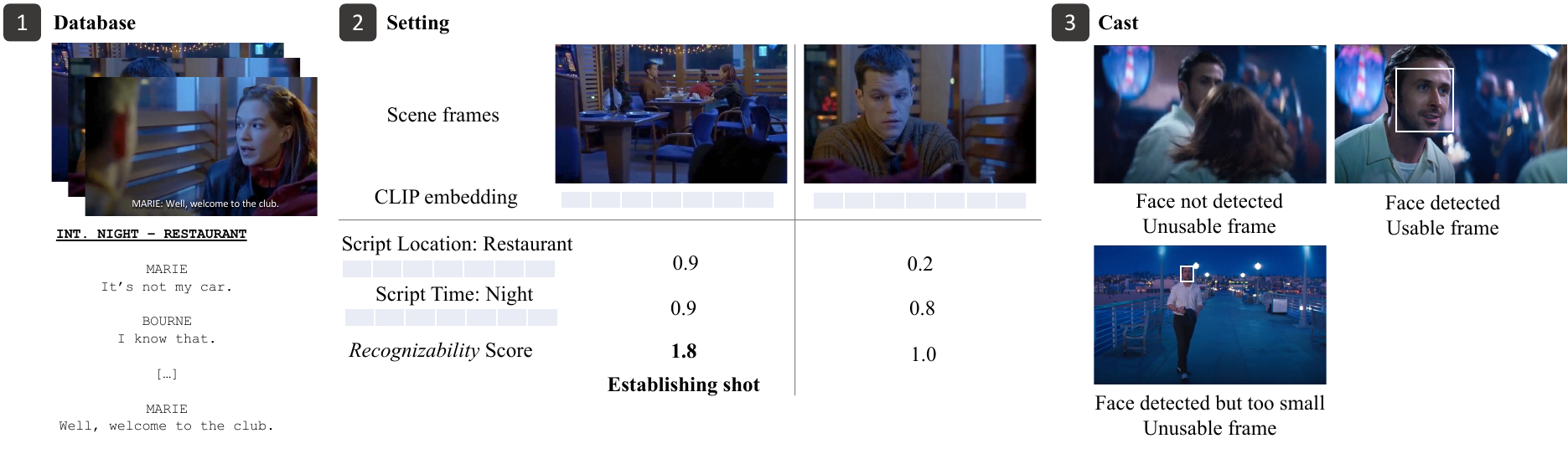}
    \vspace{-15pt}
    \caption{Preprocess database. Setting: we compute
     the CLIP~\cite{radford2021learning} visual-text similarity between setting tags and visual frames to acquire \emph{recognizability} score of setting tags for each frame.
     Cast: we detect the front face to acquire a \emph{recognizability} score for each character in each frame.
}
    \label{fig:dataset}
    \vspace{-5pt}
\end{figure*}

\vspace{-2pt}
\subsection{Preprocess Database}
To fully support the functionality of \model to choose frames for each line of dialogue,
we need frame-level annotations on 1) setting: a recognizable location, time-of-day, and 2) cast: recognizable character identity (the name of actor), gender, and age for all frames as shown in Figure~\ref{fig:dataset}.

ScriptViz builds atop MovieNet\,\cite{huang2020movienet} a database of $1,100$ films annotated 
at different levels with various attributes such as movie-level genera tags, scene-level location tags, shot-level cinematic style tags, and frame-level character bounding box annotation. 
A subset of $247$ of the MovieNet films has been annotated at the finest level containing about $300K$ scenes.
The annotations include scene boundaries, scene-level location tags $p_{loc}$, 
shot boundaries, keyframes for each shot, 
and frame-level body bounding boxes annotating an identity tag for each character $c_{ID}^i$.

\vspace{0.5em}
\noindent\textbf{Setting attributes.}
Among $\sim300K$ scenes that we used,
each is annotated with location tags from $90$ categories, such as living room, restaurant, spaceship, etc as detailed in the supplementary material. 
We use a time-of-day classifier~\cite{taha2016day} to assign binary labels ``day or night" to each frame. 
For each scene $s$,
we have its setting annotations
$$\cP(s) = \{p_{loc}, p_{time} \}.$$

\vspace{0.5em}
\noindent\textbf{Establishing shot.}
A key feature in scriptwriting is to showcase the scene setting with an establishing frame, where the specified location is clearly recognizable~\cite{jiang2021jointly}. 
Since close-up and medium shots are used regularly in movies, randomly picking a frame from a random shot in a scene may produce a frame where recognizable features of the location are occluded or unclear. 
Thus, we annotate each frame with a binary \emph{recognizability} tag 
indicating whether the location is recognizable in the frame.
To generate the recognizability tag,
we compute the visual CLIP~\cite{radford2021learning} embeddings for the keyframe of each shot and the textual CLIP embeddings of the location tag for the shot, which is the same as the scene-level location tag $p_{loc}$. 
We then compute the visual-text cosine similarity between the two embeddings.
We treat the similarity score (value between $0$ and $1$) as the \emph{recognizability} score of the location tag in each shot. 
The higher the score, the more visible the frame presents the location tag.
We similarly compute a \emph{recognizability} score for the time-of-day tag $p_{time}$. 
For each frame $f$ within a shot,
we have its setting annotation 
$$\cP(f) = \{p_{loc}, p_{loc_{recog}}, p_{time}, p_{time_{recog}} \}.$$
The establishing shot is the shot with its keyframe achieving the highest score on the sum of
\emph{recognizability} score on location and time-of-day among the scene.

\vspace{0.5em}
\noindent\textbf{Cast attributes.}
MovieNet only provides a body bounding box and identity annotation for cast members.
Then we run a front face detection\,\cite{huang2018person} to obtain a binary annotation for each character's facial bounding box indicating whether the front of their face is visible. 
We further measure each character's recognizability in a frame according to the ratio of their body bounding box and the size of the frame, and front face visibility.
If the front face is visible and its body bounding box ratio is over $10\%$ of the frame size, its binary recognizability score is $1$,
which indicates this character can be easily recognized from the frame.

As each character in the movie is associated with a cast in real life, we use \citet{imdb} to crawl information about a cast on gender, age, country, etc, when he/she stars in the movie. 
For each scene $s$, 
we have its $i$-th cast's annotation 
$$\cC(s) = \{c^i_{ID}, c^i_{gender}, c^i_{age} \}.$$
For each frame $f$ within a shot,
we have its $i$-th cast's annotation 
$$\cC(f) = \{c^i_{ID}, c^i_{recog}, 
 c^i_{gender}, c^i_{age} \}.$$

\noindent\textbf{Meta information.}
Movies include rich meta information that can provide valuable context and support for script visualization, helping scriptwriters better search for what they want. 
The movie's title, production year, and genre can give an initial idea of what to expect, helping the scriptwriters anticipate the tone and style of the film's visuals.
We associate scenes with the movie they are coming from and assign movie-level annotation to these scenes,
$$\cM(s) = \{s_{year}, s_{genre}, s_{title}\}.$$

\begin{figure*}[!t]    
    \includegraphics[width=0.99\linewidth]{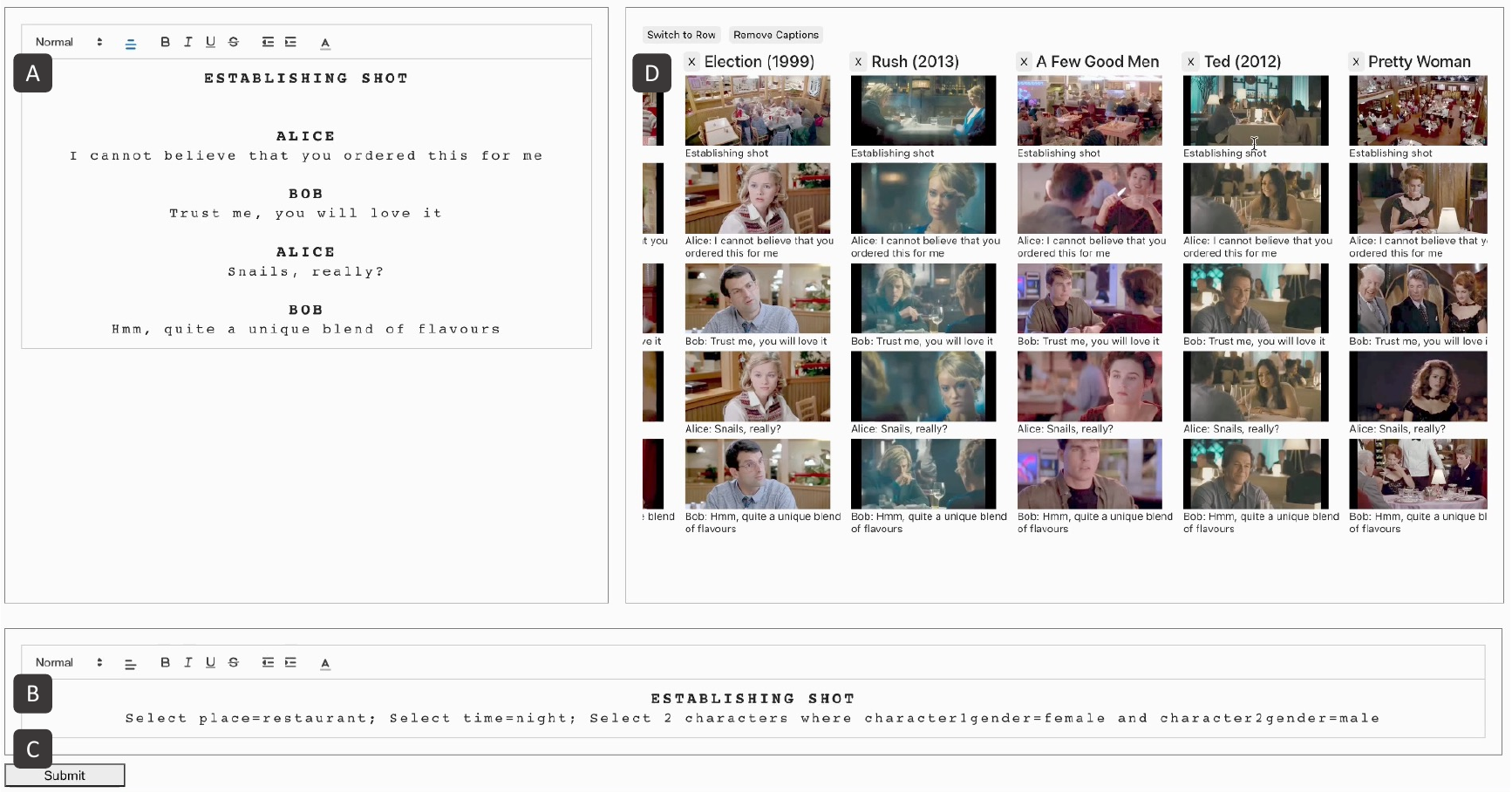}
    \vspace{-5pt}
    \caption{ScriptViz interface that consists of four components. 
    Users type in scripts (A) in AMPAS standard, add attributes control (B) component in SQL format, and click the submit button (C). 
    ScriptViz provides visualized outputs (D) from different movie scenes, where each containing one established shot overview and images for each line of dialogue in the script.
	}
    \label{fig:ui}
    \vspace{-5pt}
\end{figure*}

\vspace{-2pt}
\subsection{Search Scenes from Database}
\label{subsec:scene}
The system formulates queries to the database that take user-defined  \emph{fixed} attributes and  \emph{variable} attributes. 
These queries on fixed attributes aim to find scenes that fulfill the specified criteria, ensuring relevance and coherence with the script's narrative. 
Meanwhile, queries on variable attributes are designed to offer wide options, stimulating writers' creativity and thought processes.

To allow flexibility in querying the database and provide desired visualization,
we use structured query language (SQL)~\cite{jamison2003structured}.
SQL enables users to specify the attributes that are fixed in their minds and present variance on attributes that are not sure.  
Exemplar descriptions are like this:
\emph{select Place=Bedroom where MovieYear>1980, Time-of-day=Variable, Character1Gender=Female where Character1Age>40 and Character2=Jean}.
The user input can be denoted as $$\{\cP_{user}, \cC_{user},\cM_{user} \}.$$
%

The scene search starts from fixed movie-level attributes specified by writers,
and keeps movies satisfying the tags, such as movie genre and year.
For variable attributes, the system samples movies with distinct values for the variable attributes.

Then, the method comes to the scene level. 
1) For fixed location attributes, if the user's input belongs to the location tag category, it directly keeps scenes with the specified tag. 
If the user's input does not belong to the existing category,
we compute the text-visual cosine similarity between the textual CLIP embedding of user input and the visual CLIP embedding extracted from the establishing shot's keyframe of each scene.
\model returns the scenes in the order of their similarity score. 
2) For variable location attributes,
the system averagely samples scenes from each location value in the location category list.
3) For character attributes, we keep scenes with character numbers larger than the writer's specification on fixed attributes, and each gender/age number is enough if users specify gender/age. 
If users do not specify the number of characters in a scene, we will automatically determine it by the script itself.

Specifically, this process finds  
a specific movie scene and assigns casts for each character in the script
$$s^\ast \quad \text{s.t.} \quad \cP(s^\ast)\in \cP_{user}, \cC(s^\ast)\in \cC_{user}, \cM(s^\ast)\in \cM_{user}.$$

\vspace{-2pt}
\subsection{Visualize Dialogue with Frames}
To further assign frames for each dialogue,
we need to assign a cast from the database to each character in the script and make sure the cast in each frame is recognizable.
The input script dialogues are formatted in line with the \citet{oscars} standard, with each line explicitly tagged with the name of the character delivering it.
The scene search process in Sec.~\ref{subsec:scene} keeps scenes where the character number is enough for user-specified fixed character attributes. 
Since the character number is relatively small, usually less than 5,
we enumerate all possible cast choices here to figure out possible cast combinations to assign frames to the dialogue.

According to the writer's SQL input, \eg
\emph{Character1Gender=Female, where Character1Age>40, and Character2=Jean},
we first assign a cast that satisfies $1$-th character's fixed attributes and then do the next.
As for the variable character attributes, our system enumerates all the possible casts.
This process assigns a cast from the movie database that matches the name, gender, or age depending on users' choices on fixed and variable attributes.

To find a frame for a line in the script 
associated with $m$-th cast from the movie scene in the database,
we find a $f^\ast\in s^\ast$ from the movie scene that contains the corresponding cast with a recognizable front face,
$$f^\ast \quad \text{s.t.} \quad \cC(f^\ast) = \{c^m_{ID}, c^m_{recog}=1, 
 c^m_{gender}\in\cC_{user}, c^m_{age}\in\cC_{user} \}.$$

\begin{figure*}[!t]    
    \includegraphics[width=0.99\linewidth]{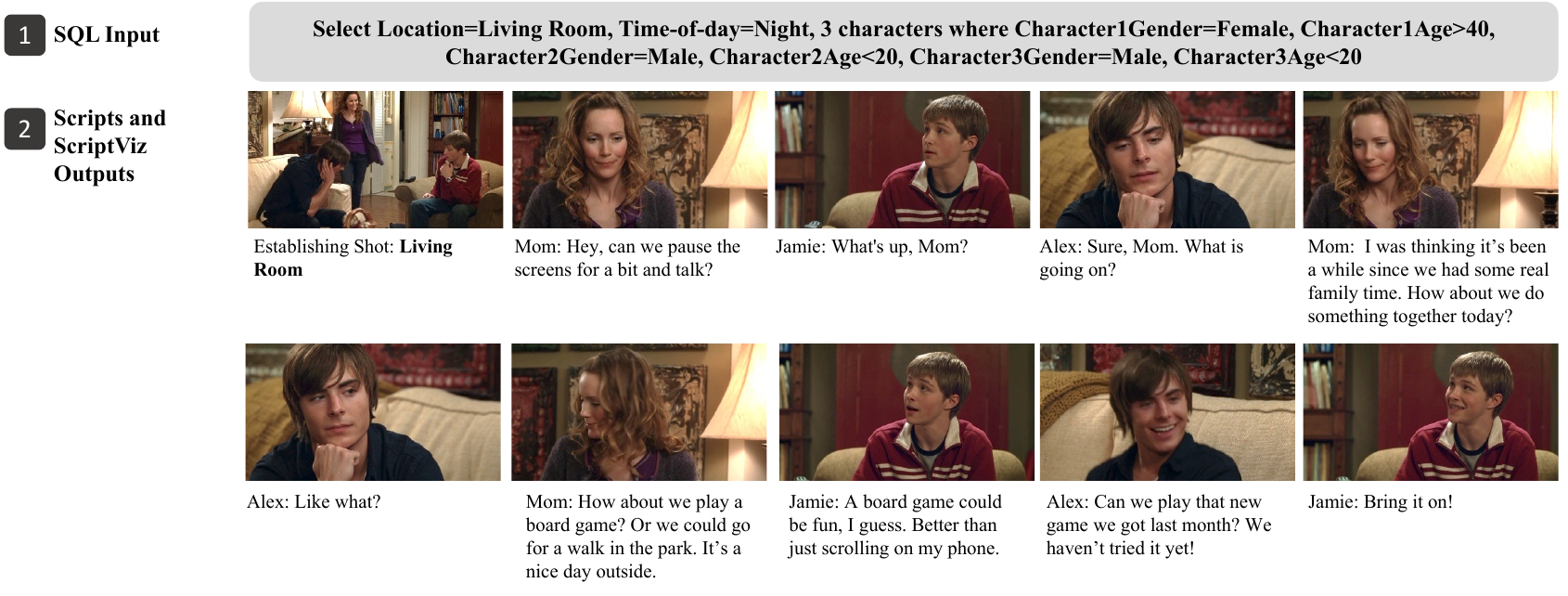}
    \vspace{-15pt}
    \caption{
	A scriptwriter writes a script containing three characters and nine lines. 
    ScriptViz returns one overview of the established shot and nine images showing three characters.
    }
    \label{fig:multichar}
    \vspace{-5pt}
\end{figure*}

\vspace{-2pt}
\subsection{Interface}
As shown in Figure~\ref{fig:ui}, the scriptwriter can modify the script in real time, compare images, and select/modify individual images to support dialogue-aligned visualization that engages scriptwriters. For each output, all images come from one movie scene to ensure visualization consistency.

\noindent\textbf{Input and output.} 
Users enter the script text (\textbf{A}) and an SQL statement defining the fixed and variable attributes (\textbf{B}). 
After clicking the submit button (\textbf{C}),
ScriptViz produces several outputs from different movies, each containing one overview and images for each line of dialogue in the script (\textbf{D}). 

While the SQL defines the setting, character number, gender, age, etc, the script determines the number of visualized images, as shown in Figure~\ref{fig:multichar}.
For each setting or character attribute, we provide a list of possible fixed attributes, e.g. male or female for character gender, day or night for time of day, etc. As for location, we provide a list of commonly used location tags in our supplementary PDF and allow free text input via CLIP-based embedding search, e.g. on the weather.

\noindent\textbf{Modification.} 
Users can select alternative images for each line from the same movie scene if they do not like the default choice by clicking the image in \textbf{D}. 
Users can also modify the script or the SQL statement and resubmit to get new images. ScriptViz produces new images in real time after the user clicks the submit button (\textbf{C}) in the interface.

\begin{figure*}[!t]    
    \includegraphics[width=0.99\linewidth]{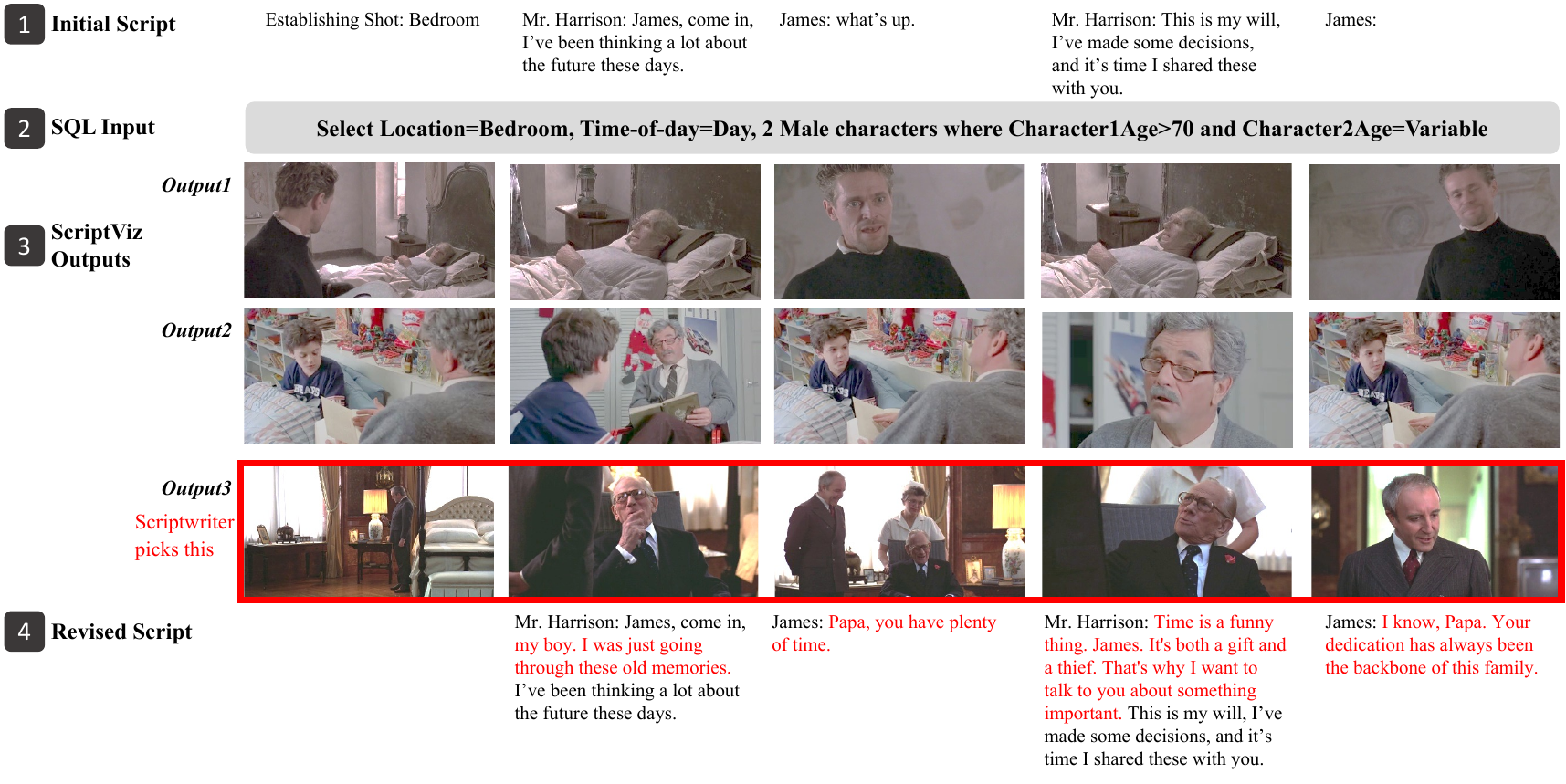}
    \vspace{-13pt}
    \caption{A screenwriter types in a script and selects their story's fixed and variable attributes. \model retrieves sequences for the input script. The screenwriter iterates on the script based on the proposed sequences.
    The visualization helps to enrich the details of the existing dialogue and write the unfinished dialogue. 
}
    \label{fig:fill1}
    \vspace{-7pt}
\end{figure*}

\section{Usage Scenarios}
\label{sec:results}

Scriptwriters input their script, set specific fixed attributes of their story, and set variable attributes or just leave variable attributes unspecified.
\model then fetches visual sequences that correspond with the entered script. 
Following these visualizations, the scriptwriter makes revisions to the script.
This reflective process is central to ScriptViz's design. By offering scriptwriters the flexibility to partially shape their script's visual aspect, ScriptViz encourages a more iterative and reflective writing process. Writers can continuously alternate between crafting their narrative and reviewing visual interpretations, ensuring that each element of the script - from character attributes to scene settings - is consciously defined to support the effectiveness of their narrative.

\begin{figure*}[!t]    
    \includegraphics[width=0.99\linewidth]{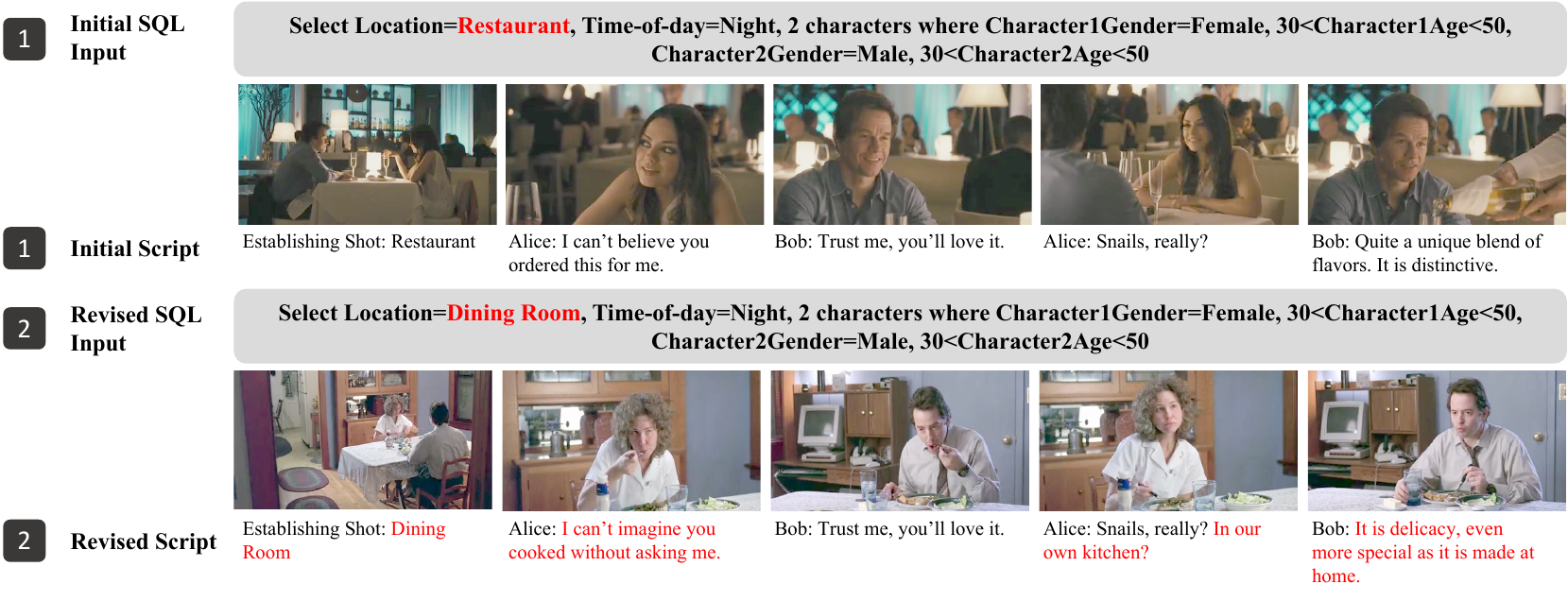}
    \vspace{-15pt}
    \caption{Comparison of visualization via changing the location attributes from ``Restaurant" to ``Dining Room" in the SQL input.}
    \label{fig:change1}
    \vspace{-5pt}
\end{figure*}

\begin{figure*}[!t]    
    \includegraphics[width=0.99\linewidth]{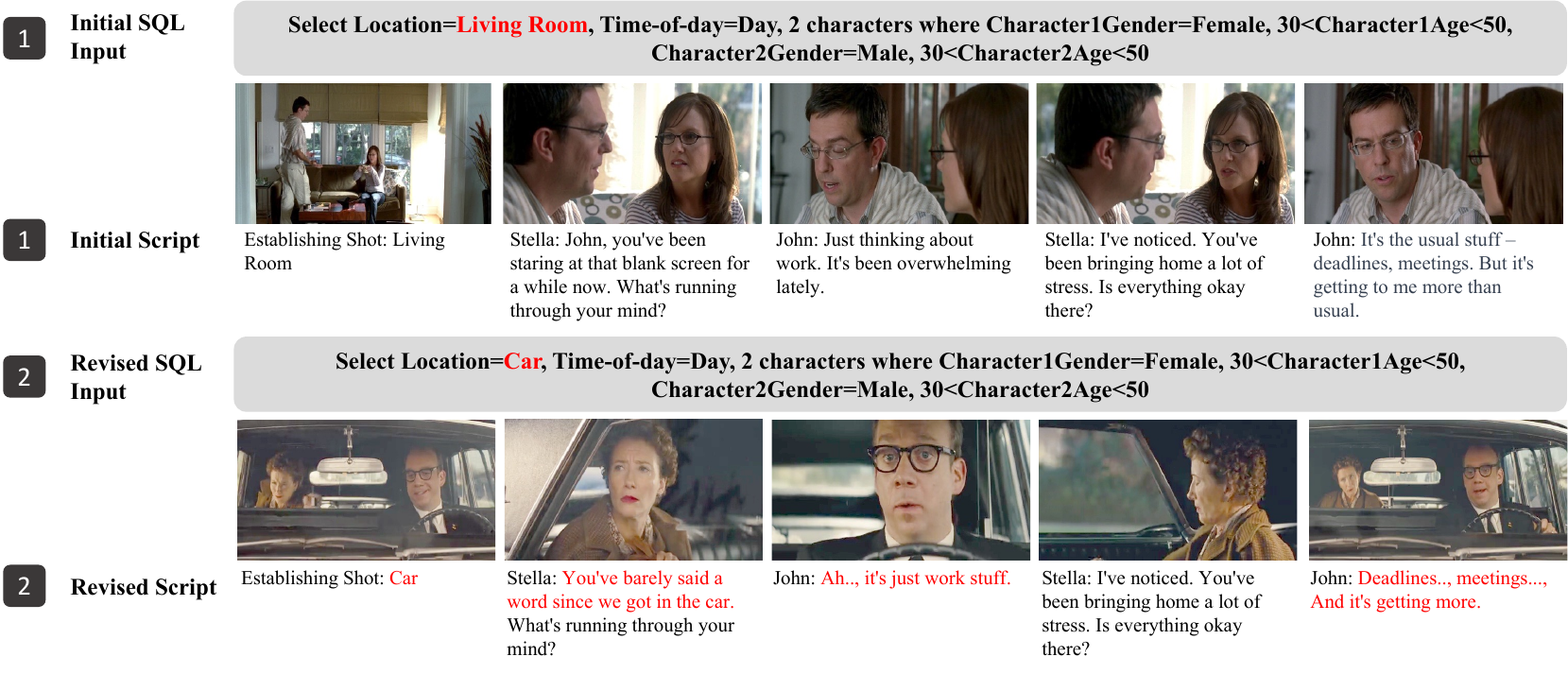}
    \vspace{-15pt}
    \caption{Comparison of visualization via changing the location attributes from ``Living Room" to ``Car" in the SQL input.}
    \label{fig:change2}
    \vspace{-5pt}
\end{figure*}

We showcase three additional practical usage scenarios from our user study in the following two typical categories.
1) See variable attributes to establish settings in 
Figure~\ref{fig:fill1} and our teaser Figure~\ref{fig:teaser}. 
2) Change fixed attributes to revise scripts in  Figure~\ref{fig:change1} and Figure~\ref{fig:change2}.

\vspace{0.5em}
\noindent\textbf{See variable attributes to establish settings.}
To exemplify how \model integrates into the scriptwriting process, we extract the writing samples made by professional scriptwriters from our user study.
A scriptwriter (P2) writes a ``will discussion" scene in a bedroom between Mr. Harrison: An elderly man in his late 80s, wise, with a warm and kind demeanor and James: A younger man, Mr. Harrison's son or grandson, thoughtful and attentive. 
Before using \model, the scriptwriter already has some basic dialogue in his mind, as shown at the top of Figure~\ref{fig:fill1}.
By specifying the fixed location attribute ``Bedroom", and character attributes ``2 male characters where Character1Age>70 and Character2Age=Variable",
\model returns results in Figure~\ref{fig:fill1}, where the age of James is 30 in the first row, 10 in the second row, and 50 in the third row.  
Among all these visualizations, the scriptwriter chooses the third one as it better matches James's thoughtful and attentive personality.
Based on the third visualization of an old man talking with a younger man, 
their facial expressions, and the inviting atmosphere, 
the scriptwriter slows the pace of dialogue with longer sentences and words like ``my boy" and ``plenty of time".
This slow pace better shows Mr.s Harrison's personality with a warm and kind demeanor.

\vspace{0.5em}
\noindent\textbf{Change fixed attributes to revise scripts.}
A scriptwriter (P5)  writes an argument scene between a couple Alice and Bob,
as shown in Figure~\ref{fig:change1}.
Initially, she sets the scene at nighttime in a ``restaurant".
She wants to write Bob, who is characterized as confident and a bit presumptuous, as shown by his interaction with Alice. 
However, she is still not satisfied with the atmosphere of this conflict.
She then realized that as Alice and Bob already know each other, it might be more appropriate to set up this conflict in a more intimate or casual setting, like home.
She changes it to a nighttime ``dining room".
Accordingly, she changed the dialogue of the first sentence from ``order food" to ``cook food" as a metaphor for Bob invading Alice's kitchen. 
In the third sentence, 
Alice directly expresses her unhappy feelings by stressing the ``kitchen".

As shown in Figure~\ref{fig:change2}, another scriptwriter (P8) first creates a dialogue in a ``living room" between John and Stella to establish the tense situation John faces at work.
After changing the location to a moving ``car", 
John is busier, with his attention divided between the road and the conversation, adding more tension to his situation.
Accordingly, John speaks shorter in fits and starts. 
In the car scene, John and Stella are positioned in a linear arrangement, with John focused on driving and Stella as a passenger. This creates a sense of separation.
\section{User Evaluation}
\label{sec:experiment}

\subsection{Setting}
We conducted a study to evaluate the effectiveness of ScriptViz in aiding scriptwriting. 
The study consisted of three parts: 1) a pre-experiment survey and demo, 2) scriptwriting, and 3) a questionnaire, rating, and interview.

It involved 15 participants with detailed information in Table~\ref{table:userinfo}, including 8 professionals (P1-P8) who worked in the film industry for years and 7 hobbyists (H1-H7) who graduated from scriptwriting majors and wrote scripts in their spare time with over twenty script creation experiences.
We conducted remote studies on Zoom. Participants used Chrome Remote Desktop to access the system on the interviewer’s computer.

The system runs on a M1 MacBook Pro Laptop with 16GB of memory. 
With the processed database, it usually takes 1s$\sim$5s to get a visualization of twenty variant results.
The size of the database, its annotations, and preprocessed embeddings take 210GB and contain 300K scenes. In the preprocessing, we utilize an i7 desktop with a GPU 4090 to compute the CLIP image embedding for the database, which takes approximately 6 hours.

\begin{table}[t!]
    \centering
     \caption{Information about the participants of the system evaluation.}
     \label{table:userinfo}
     \vspace{-8pt}
     \resizebox{\columnwidth}{!}{
    \begin{tabular}{cccc}
    \hline
    Participant & Experience & Creation Frequency & Fields\\
    \hline
    P1 & Professional (8 years) & 3$\sim$4/week & Film-making, Animation\\
    P2 & Professional (5 years) & Everyday & Film-making\\
    P3 & Professional (7 years)& Everyday & Film-making \\
    P4 & Professional (6 years) & Everyday & Animation \\
    P5 & Professional (7 years)& 1$\sim$3/week & Film-making \\
    P6 & Professional (9 years)& 2$\sim$3/week & Film-making \\
    P7 & Professional (6 years)& 1$\sim$3/week & Short Film-making \\
    P8 & Professional (8 years)& 3$\sim$4/week & Short Film-making \\\hline
    H1 & Hobbyist (5 years)& 1$\sim$2/month & Theater \\
    H2 & Hobbyist (6 years)& 1$\sim$2/month & Gaming \\
    H3 & Hobbyist (4 years)& 4$\sim$5/month & Advertisement \\
    H4 & Hobbyist (3 years)& 3$\sim$5/month & Vlogger \\
    H5 & Hobbyist (6 years)& 3$\sim$4/month & Advertisement \\
    H6 & Hobbyist (6 years)& 1$\sim$4/month & Advertisement \\
    H7 & Hobbyist (4 years)& 2$\sim$5/month & Advertisement \\
    \hline
    \end{tabular}
     \vspace{-8pt}
    }
\end{table}

\vspace{4pt}
\noindent\textbf{Pre-Experiment.}
Before the study, we conducted a pre-experiment survey to gather information about the participants’ experience with scriptwriting, as shown in Table~\ref{table:userinfo}. The survey included questions about the number of years they have been writing scripts, how often they write, what they write for, and their background.

\vspace{4pt}
\noindent\textbf{Baselines.}
Participants were then asked to complete three tasks that involved writing a scene from scratch. 
They were given 2 hours to complete three tasks, 
where each task involved writing different scenes they want with four methods:
1) \emph{text-only} writing without any visualization. It is the status quo approach scriptwriters commonly use and we allow users search for images, sketches, and videos using Google;
2) writing with \emph{NoAlign} baseline: to verify the design of dialogue-aligned visualization, 
we set this baseline visualization as "retrieving one establishing shot and one picture for each character"; 
3) writing with \emph{NoRecog} baseline: to verify our design on database preprocessing that supports fixed and variable attributes query, we remove the frame-level recognizability of our tags and retrieve any frame that satisfies the scene-level tags and character body bounding boxes; 
4) writing with \model.

All the above methods use the same text entry/script writing interface as our tool, only differing in the images shown in the visualization box (text-only leaves this image area blank). Thus, after training users in the ScriptViz interface, they are familiar with all of the baselines as well. It won’t introduce additional bias in the usage of different baselines.
Participants were given a 30-minute demo of \model to familiarize themselves with the tool. 
We also record the time when they confirmed that their scriptwriting reached their expectation in Table~\ref{tab:time}.

The order of the four methods is shuffled to address ordering effects. 
After completing the task, participants were asked to complete a questionnaire that included four aspects on writing support, system performance, play out, and visual addition.
Participants rated their agreement with each statement on a Likert scale ranging from 1 (Strongly Disagree) to 5 (Strongly Agree). The questionnaire took approximately five minutes to complete.
The results are shown in Figure~\ref{fig:user}, noting the text-only baseline is not applicable to these questions.

\begin{table}[!t]
\caption{Time comparison of using different approaches to write scripts that reach writers' expectations.}
\label{tab:time}
\vspace{-8pt}
\resizebox{0.99\columnwidth}{!}{
\begin{tabular}{l|cccc}
\toprule
dialogue length/sentence              & ScriptViz & NoAlign & NoRecog & text-only  \\ \midrule
5-10 &  7.6$\pm$2.3 min  & 12.2$\pm$3.2 min  &  18.4$\pm$3.8  min & 19.6$\pm$4.3 min     \\
11-15     &  13.7$\pm$2.7 min & 16.0$\pm$3.7 min  &  19.7$\pm$4.4  min  &   24.2$\pm$5.1 min  \\
16-20     &  17.7$\pm$3.3 min  & 19.0$\pm$4.8 min  &  21.7$\pm$4.3  min  &   27.3$\pm$5.3 min  \\
\bottomrule
\end{tabular}
}
\vspace{-8pt}
\end{table}

\subsection{Results}
\noindent\textbf{\model.}
All 15 participants suggested our visualization tool helped with their script writing, and 12 out of 15 strongly agreed with this (Q1-1).
The participants unanimously noted ScriptViz's capacity to introduce new ideas into their scripts (Q1-2).
14 participants found the retargeted sequences provided by the system were relevant to their script’s scenes (Q2-1). 
The system helped them explore different visual possibilities.
All the participants agreed that the system made it easy to imagine what my script could look like (Q3-1).
They unanimously agreed that the tool facilitated the exploration of diverse visual possibilities, making the scriptwriting process more immersive and enhancing the visual dimension of their work (Q3-2).
P3 liked the way that "the tool breathed life into my script, allowing me to visualize scenes in ways I hadn't considered before."
Importantly, participants indicated that using ScriptViz helped them develop the visual dimension of scripts (Q4-1) and influenced their thinking about visual screenwriting, which facilitates downstream filming. 
"It's like having a visual collaborator guiding the creative process." (P1) These affirm ScriptViz's role as a valuable tool for scriptwriters, fostering creativity and elevating the visual elements of their work.

\begin{figure}[!t]
    \includegraphics[width=0.99\columnwidth]{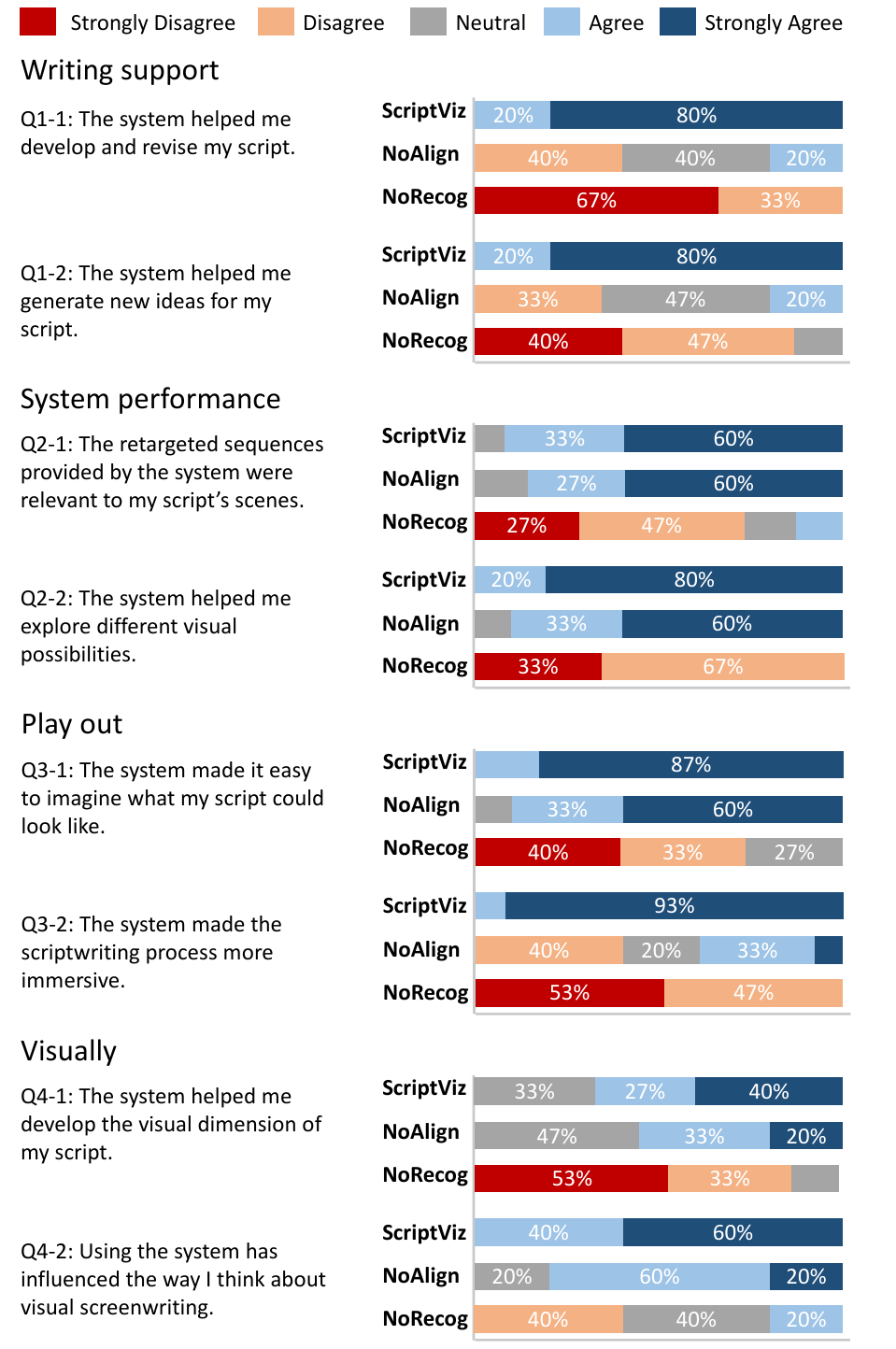}
    \vspace{-10pt}
    \caption{Questionnaire results.}
    \label{fig:user}
    \vspace{-5pt}
\end{figure}

\vspace{4pt}
\noindent\textbf{Baselines.}
Table~\ref{tab:time} indicates that ScriptViz can help scriptwriters achieve their desired scripts faster than others.
The \textit{NoAlgin} baseline 
is able to retrieve the right images for the setting and each character but is not aligned with the script, making it harder for writers to build connections between scripts and visuals.
It leads to lower scores in writing support in Figure~\ref{fig:user} and slower speed to achieve desired scripts, which underscores the importance of our dialogue-aligned design. 
The \textit{NoRecog} baseline encounters issues in retrieving the right images corresponding to users' input, which results in a much lower score in all aspects in Figure~\ref{fig:user} and a longer time to finish the creation.
This proves the effectiveness of our attributes recognizability preprocessing design for script visualization.
The \textit{text-only} baseline uses Google search, but the content may not be able to provide the desired visualization to scriptwriters and requires them to mentally visualize the scripts. This results in the longest time spent in scriptwriting.

\begin{figure}[!t]    
    \includegraphics[width=0.99\linewidth]{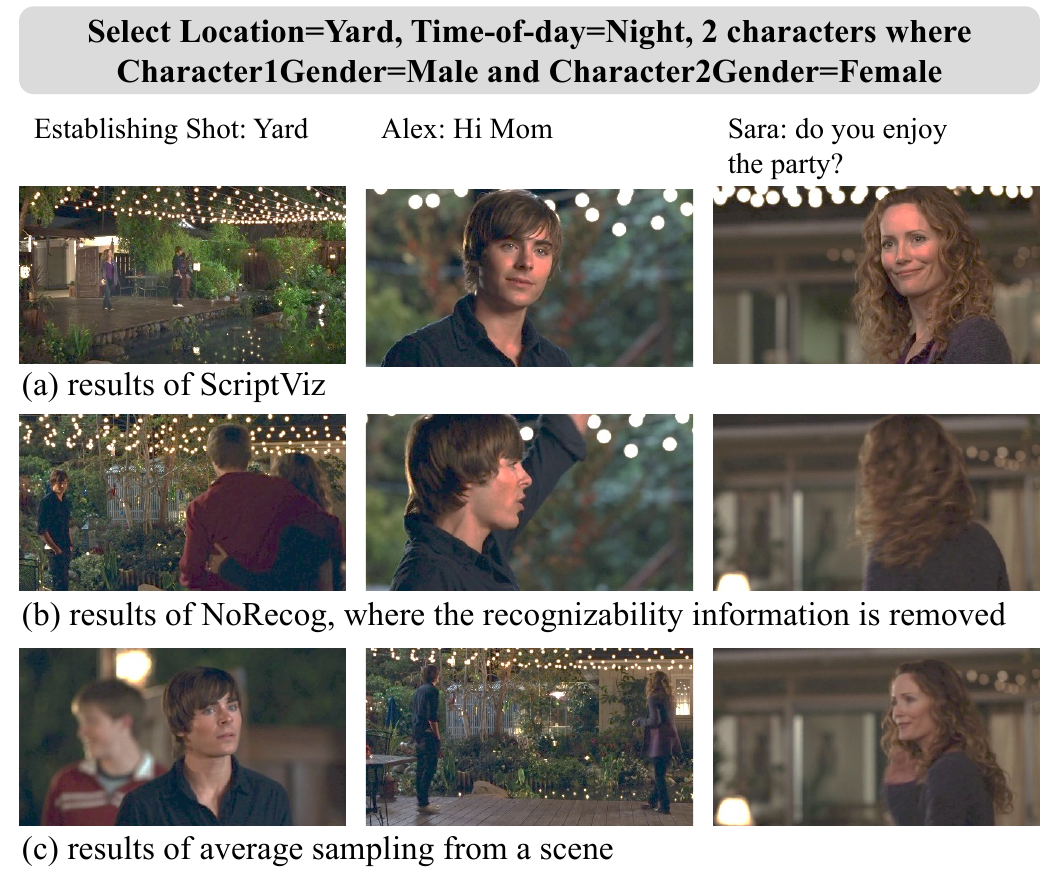}
    \vspace{-13pt}
    \caption{Qualitative clarity comparison of different approaches.
    }
    \label{fig:clarity}
    \vspace{-4pt}
\end{figure}

\vspace{4pt}
\noindent\textbf{Clarity of retrieved frames.}
To verify the effectiveness of our data preprocessing, 
we additionally visualized 1,000 samples (an example is shown in Figure~\ref{fig:clarity}) of (a) ScriptViz, (b) NoRecog,
(c) Average sampling frames from the scene,
and asked our 15 users to rate its binary clarity.
The whole samples are evenly distributed to each user, and each is given 200 samples of three visualizations to ensure each sample is rated by 3 different users.
For the cast attributes, if any of the cast is not clear to see, the user rate 0; otherwise, 1.
If the setting attributes are unclear, the user rates 0; otherwise, 1.

Table~\ref{tab:clarity} and Figure~\ref{fig:clarity} indicate that 
our pre-processing method can successfully retrieve clear frames
of establishing shots and characters.

\begin{figure*}[!t]    
    \includegraphics[width=0.99\linewidth]{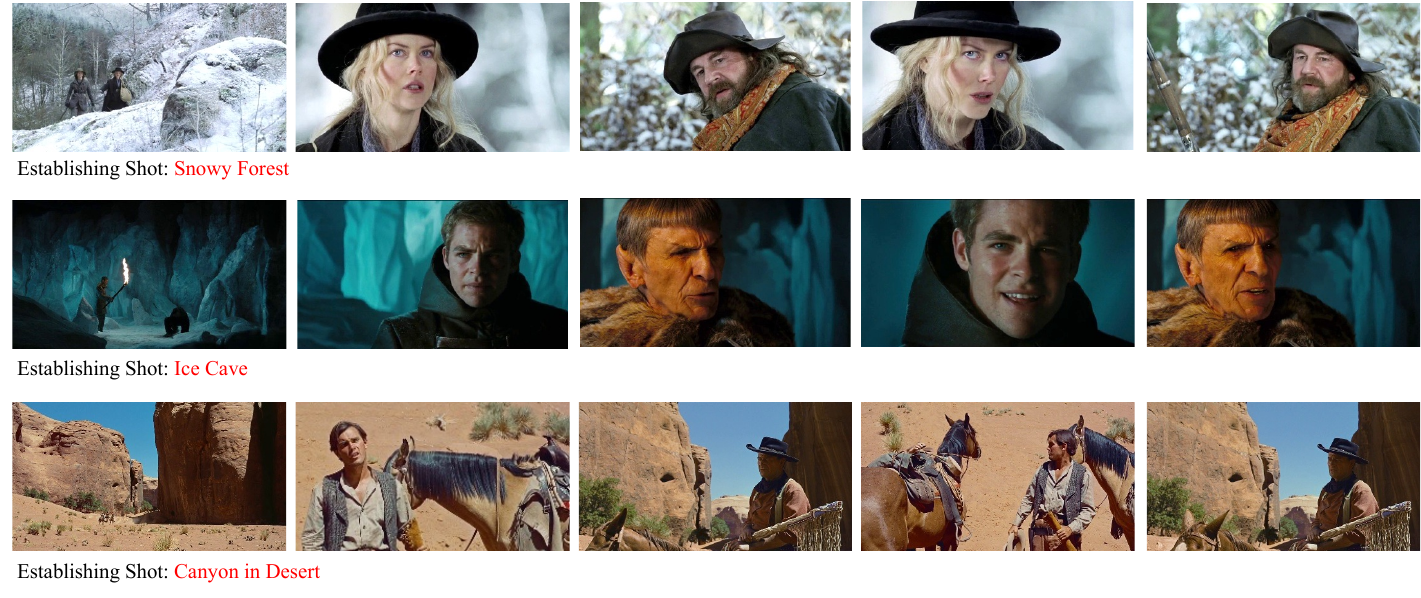}
    \vspace{-12pt}
    \caption{
    The visualization results in the location value outside the category list in MovieNet 
    via computing the text-visual similarity between the textual CLIP embedding of user input and the visual CLIP embedding extracted from the establishing shot’s keyframe of each scene.
    Noting that the ``desert" in the teaser image does not belong to the existing category list, we show more results here, such as ``Snowy Forest," ``Ice Cave," and ``Canyon in Desert."}
    \label{fig:varyloc}
    \vspace{-4pt}
\end{figure*}

\vspace{4pt}
\noindent\textbf{Richness of retrieved sequences.}
Figure~\ref{fig:varyloc} shows that
\model is able to provide diverse visualization on location beyond the tag list provided by MovieNet, which greatly enlarges the supported locations.
We additionally asked our 15 users to input 100 of their commonly used place tags and rate 1 if the returning establishing shot satisfies them; otherwise 0.
The satisfying rate is 94.6\%.
This shows the effectiveness of our text-visual similarity computation in Sec.~\ref{subsec:scene} to support place tags outside the existing category.

\begin{table}[!t]
\caption{Quantitative clarity comparison of different approaches on 1K samples.}
\label{tab:clarity}
\vspace{-8pt}
\resizebox{0.75\columnwidth}{!}{
\begin{tabular}{l|ccc}
\toprule
item              & ScriptViz & NoRecog & Average sampling  \\ \midrule
Setting &  97.1\%  & 30.2\%  &  6.5\% \\
Cast   &  97.7\% & 36.8\%  &  7.4\%   \\
\bottomrule
\end{tabular}
}
\vspace{-8pt}
\end{table}

\subsection{Semi-Structured Interview}
Finally, participants were interviewed for 30 minutes using a semi-structured format to understand why \model helps in the scriptwriting process.  
These interviews aimed at gaining a deeper understanding of their creative processes during scriptwriting and their experiences with existing tools.
We took notes during our semi-structured interview with users’ consent. Two researchers analyzed them to compare and identify commonalities or discrepancies.

\vspace{4pt}
\noindent\textbf{Why do we need ScriptViz visualization?}
Scriptwriters have found ScriptViz to be a transformative tool in their creative process. 
The tool offers a unique approach to visualizing scenes, providing immediate visual references based on script text and dialogue. 
Compared to baselines, it is more frequent for participants to ``see details in the visualized frames that help to develop and revise scripts" (P2, P3, P5, P6, P7, U3, U4, U5, U6).
This capability ``sparks fresh ideas" (P4), ``helping writers flesh out details" (P7), and ``explore alternative paths" (P8) in their storylines. 
ScriptViz allows writers to mentally immerse themselves in the moment, making it easier to visualize character movements, facial expressions, and overall scene dynamics.

Compared to existing methods like internet image search (similar to our \textit{text-only} baseline) or moodboards (similar to our \textit{NoAlign} baseline), 
ScriptViz is deemed more accurate and efficient (P1, P3, P4, P5, P6, P8, U1, U2, U5). 
It provides more relevant and fitting visual references or suggestions for scenes described in the script and streamlines the process of finding and integrating visual elements with the script. 
By being more accurate, ScriptViz likely presents options that better align with the envisioned scenes, reducing the need for extensive manual searching or curation. Moreover, its efficiency implies that it saves time and effort for the scriptwriters compared to methods like internet image searches or moodboards, which may require more manual effort in compiling and organizing visual references. 
``The visuals provided by \model help me better focus on my writing comparing to others" (P5)
``The tool is much more efficient than my previous daily life workflow where I need to watch movies and find keyframes" (P8)
All participants also agreed that the tool is easy to be integrated seamlessly into the writing process, offering a dynamic visual dimension that enhances creative endeavors.

\vspace{4pt}
\noindent\textbf{Is the system design reasonable?}
One of the standout features is the consistent yet diverse visual possibilities provided by fixed and variable attributes control (P1, P2, P3, P5, P6, P7, U1, U2, U3, U6). 
"ScriptViz doesn’t just offer a single interpretation of my script; it provides a spectrum of visuals, allowing me to explore different cinematic approaches while maintaining narrative coherence." (P2)
Writers primarily focus on visualizing key elements of characters and settings, aiming to convey the essence of the moment through visuals effectively (P1, P2, P4, P7, P8, U2, U5, U7). 
But P3 also commented ``it also depends on the key elements in my script." 
It is helpful to have more attributes to customize and refine the visualizations. ``It would allow for even greater flexibility and creativity in bringing my script to life visually." (U3)

The visualizations provided by ScriptViz were generally in line with scriptwriters' expectations and ``helped me visualize my scenes more effectively." (P3)
Some participants (P1, U2) also found that this tool is more helpful in the starting stage of their script creation, and the information in the visualization gets saturated when there are more than 7-8 frames.
The illustrations do not necessarily need to be exact replicas of what the scriptwriters imagined. Instead, they should aim to capture the essence and mood of the scenes as described in the script. The illustrations should provide relevant visual references that align with the overall vision and tone of the script, helping to enhance the storytelling process (P1, P2, P3, P5, P7, P8, U1, U3, U7).
Using existing movies as a reference can profoundly offer inspiration and reference points for visual elements, 
but familiarity with the movies used in ScriptViz's database is not a prerequisite for utilizing the tool (P1, P4, P5, P6, P7, U3, U4, U5, U6, U7).
P4 said ``the tool provided visual references based on the script text and dialogue, regardless of whether I knew the source movies."

\vspace{4pt}
\noindent\textbf{Summary.}
Overall, ScriptViz has the potential to contribute to the broader understanding of visual storytelling in scriptwriting. 
By providing writers with a smart platform for visual feedback, ScriptViz can help writers experiment and test their screenplays, leading to more engaging and creative scripts.

\section{Discussion}
\label{sec:limitations}

\noindent\textbf{Failure cases.}
ScriptViz relies on the quality of the movie database.
There are some wrong annotations coming from manual processes or automatic annotations in the database, e.g., the cast identity, the front face of the case, and the place tag. 
The percentage of such wrong annotations is tiny according to our study in Table~\ref{tab:clarity}. There will be failure cases like the system cannot find the right place, or find a wrong cast for the character in the scripts, or the visualized frames not showing the faces. Since \model returns multiple outputs and these failure cases are rare, they do not affect the practical usage of ScriptViz.

ScriptViz also relies on the availability of the database, which may not cover all the genres, themes, and styles of the scripts. 
For example, less mainstream genres may be underrepresented, limiting the tool's ability to accurately visualize scripts from these categories. To address this, future efforts could focus on expanding and refining the movie database, incorporating a wider range of films across various genres and styles. 
Our proposed database preprocessing techniques can help to extend to more data.

\vspace{4pt}
\noindent\textbf{Experiment limitations.}
While users expressed enthusiasm for the tool, it's important to acknowledge some experimental limitations.
The study's scope was constrained to a short session, which doesn't capture the extensive duration—spanning months to years—that scriptwriting typically involves. Participants were instructed to start new stories, yet having a more developed narrative could significantly affect how visualizations are utilized. Additionally, scriptwriting often unfolds as a collaborative effort, suggesting that individual preferences may impact perceptions of specific visualizations. To collect insights that more accurately reflect the scriptwriting process, conducting a randomized, controlled, and longitudinal field trial would be invaluable. Such a study, engaging story writers over a longer period, would offer a deeper understanding of how ScriptViz is perceived and incorporated into diverse creative workflows, providing a clearer picture of its utility across various writing scenarios.

\vspace{4pt}
\noindent\textbf{Visualization controllability.}
To allow more controllability and flexiablity to the final visualizations, we can pinpoint two avenues for improvement. 
Firstly, to better suit scriptwriters' narrative goals, we could introduce sequence editing features, such as montage idioms~\cite{leake2017computational} for high-level rearrangement or frame manual selection and addition/removal, which would let users craft montages that better reflect their intentions. 
Secondly, we could expand the tool's generative capabilities to address the data limitations and offer more visualization exploration on top of liked sequences. Such changes could be supported with models such as InstructP2P ~\cite{brooks2023instructpix2pix} and ENJOY ~\cite{zhuang2021enjoy}. We also showcase these adjustments in the supplementary.

\vspace{4pt}
\noindent\textbf{Scriptwriting in general.}
ScriptViz visualizes one scene written by users with one scene from the MovieNet database. For each individual scene, the characters' appearance is consistent as the characters come from the same movie scene, as shown in Figure~\ref{fig:pipeline}. 
When a script has multiple scenes, the scriptwriters can specify the MovieName as fixed attributes to impose consistency on characters' appearances across the whole script. ScriptViz's capability to visualize individual scenes in various settings can help to enhance the flow and structure and serve for downstream editing and production~\cite{rao2022shoot360,rao2022coarse,rao2022temporal,ma2023automated}. 

\vspace{4pt}
\noindent\textbf{Beyond sequence of frames.}
As a future direction, exploring the incorporation of other media forms could also enhance our task of supporting visual screenwriting, enabling more immersive play-throughs and embodiment, and raising additional considerations over the final cinematic medium. Integrating audio elements, for instance, could offer scriptwriters not just a visual but an auditory sense of their scenes, enriching the narrative ambiance with music, dialogue tones, and sound effects that mirror the final production. Similarly, animations or video extracts could bring dynamic movement to the visualizations, allowing writers to observe character actions, camera movements, or scene transitions in a more lifelike manner. These multimedia enhancements would not only deepen the screenwriting process but also provide a more holistic preview of how scripts translate into the cinematic medium. Investigating the effectiveness and user reception of these features could open up new avenues for making screenwriting a more engaging and comprehensive creative experience.

\section{Conclusion}
\label{sec:conclusion}
In this paper, we propose ScriptViz, a novel tool that provides external visualization for the screenwriting process. 
ScriptViz leverages a large movie database to retrieve reference visuals that match the scripts’ text and dialogue. 
ScriptViz offers two types of control on the visual elements: fixed and variable. Fixed elements enable writers to see exactly what they want, while variable elements allow them to see different variations of the same scene.
The user study with 15 scriptwriters shows that ScriptViz is easy to use and provides relevant and diverse visuals for scriptwriting.

\begin{acks}
This work was partially supported by the Brown Institute for Media Innovation.
\end{acks}

\bibliographystyle{ACM-Reference-Format}
\bibliography{main}
\end{document}